# Rapid Assessment of Damaged Homes in the Florida Keys after Hurricane Irma

Siyuan Xian[1,*], Kairui Feng[1], Ning Lin[1], Reza Marsooli[1], Dan Chavas[2], Jie Chen[2], Adam Hatzikyriakou[1]

[1]Department of Civil and Environmental Engineering, Princeton University
[2]Department of Earth, Atmospheric, and Planetary Sciences, Purdue University

**Abstract**

On September 10, 2017, Hurricane Irma made landfall in the Florida Keys and caused significant damage. Informed by hydrodynamic storm surge and wave modeling and post-storm satellite imagery, a rapid damage survey was soon conducted for 1600+ residential buildings in Big Pine Key and Marathon. Damage categorizations and statistical analysis reveal distinct factors governing damage at these two locations. The distance from the coast is significant for the damage in Big Pine Key, as severely damaged buildings were located near narrow waterways connected to the ocean. Building type and size are critical in Marathon, highlighted by the near-complete destruction of trailer communities there. These observations raise issues of affordability and equity that need consideration in damage recovery and rebuilding for resilience.

**Introduction**

Hurricane Irma made landfall near Cudjoe Key (lower Florida Keys) on September 10, 2017, as a Category 3 storm. Irma caused widespread damage to the Florida Keys due to storm surge and waves. Informed by hydrodynamic modeling and post-storm satellite imagery, we carried out a field survey soon after (September 21-24) the event to investigate the damage to the Keys, particularly the Big Pine Key and Marathon areas.

Post-hurricane damage studies have improved our understanding of coastal vulnerability (e.g. Xian et al., 2015 and Hatzikyriakou et al., 2015 for Hurricane Sandy; Eamon et al., 2007 and van de Lindt et al., 2007 for Hurricane Katrina; Wang et al. 2017, Shao et al. for general cases). Here, we conduct a rapid damage survey and assessment for Hurricane Irma, and we use a statistical regression approach to quantify the contribution of specific vulnerability factors to the damage. Such rapid post-event assessments can provide crucial information for implementing post-storm response measures (Lin et al., 2014; Horner et al., 2011; AL-Kanj et al., 2016). The raw and analyzed data from this study appear on DesignSafe[1], a web-based research platform of the National Science Foundation's (NSF) Natural Hazards Engineering Research Infrastructure (NHERI).

**Storm Surge and Wave Simulation**

To understand the hazard and inform the field survey, we first use the coupled hydrodynamic and wave model ADCIRC+SWAN (Dietrich et al. 2012, Marsooli and Lin 2017) to simulate the storm tide (i.e., water level) and wave height for Hurricane Irma. To simulate Irma's storm tide and wave (Figure 1), we apply the surface wind (at 10-m) and sea-level pressure fields from National Center for Environmental Prediction Final (NCEP FNL) operational global analysis data ($0.25^o$ x $0.25^o$ x 6 hours). The model results, e.g., time series in Figure 1, indicate that the model satisfactorily captures the temporal evolution and the peak values of the water levels and wave heights induced by Hurricane Irma. The model results show that the highest water levels, between 2 and 2.5 m, occurred in South/Southwest Florida. However, coastal zones in this region are predominantly uninhabited and covered by wetlands, so little loss of life or property is expected. High water levels are also estimated for the Florida Keys, especially islands located on the right side of the storm track. For example, the peak storm tide in Big Pine Key and Marathon reaches up to 2 m. The model results also show that large waves with a significant wave height of about 14 m reached a few kilometers off the Florida Keys. In contrast, wave heights off the southern and southwestern coasts of Florida were small (< 2 m).

---

[1] https://www.designsafe-ci.org/#research

**Damage Survey and Analysis**

NOAA's post-storm satellite imagery[2] provides an overview of Irma's impact. The two selected survey areas in Florida Keys, the Big Pine Key and Marathon, suffered the most severe damage, according to the satellite imagery, and experienced high water levels and wave heights, indicated by hydrodynamic modeling.

Field surveys can provide detailed information for analyzing damage mechanisms. However, traditional on-site surveys require a significant time and effort, as surveyors must walk through affected areas and photograph damaged properties. Thus, we applied a rapid survey method. Rather than walking, we drove at a speed of 10 mph throughout the affected areas, taking GPS-informed pictures from the rare side windows. Over two days, the team took 3700+ pictures for 1600+ residential buildings comprised of single family and mobile homes (e.g., trailers).

Using the collected photos and the satellite images, we categorized the damage state for each surveyed residential house. Satellite images were primarily used to assess roof damage. More detailed damage mechanisms were further evaluated from the photos. We adopted FEMA's damage state criteria used in the damage assessment study for Hurricane Sandy[3]. The categories include: No/very limited damage; Minor damage; Major damage; and Destroyed.

---

[2] https://storms.ngs.noaa.gov/storms/irma/index.html#6/28.139/-81.547
[3] https://www.arcgis.com/home/item.html?id=307dd522499d4a44a33d7296a5da5ea0

We found that the destroyed and severely damaged buildings were caused largely by hydrodynamic forces induced by storm surge/waves. For example, Fig. 2a shows that storm surge/waves completely crashed the lower part of a building in Big Pine Key. Fig. 2b shows debris from damaged trailers floating in the water in a trailer community in Marathon. The observed storm surge damage is consistent with the high surge and wave heights estimated for the two sites (Figs. 3a and 3b). The assessed damage state for each house appears in Figs. 3c and 3d. The slightly and moderately damaged buildings are 72.7% and 75% of the total surveyed building for the assessed areas in Big Pine Key and Marathon, respectively. The percentages of the destroyed buildings are 13.9% and 16.9%, respectively. In both areas, the destroyed buildings are clustered. The destroyed buildings in Big Pine Key are near the coastline and narrow waterways, a strong indication that the damage was caused mainly by hydrodynamic forces. The completely destroyed buildings in Marathon cluster in the north and middle parts of the study area. The majority of those buildings are mobile homes.

Statistical analysis confirms these general observations. We use an ordered logistic regression model to correlate the damage state with the following factors: distance from the coastline (m), building type, and building size ($m^2$). Our analysis for Big Pine Key shows that the distance from the coastline is the single significant predictor of damage state (p-value < 0.001; Table 1a), as the damage is dominated by buildings located near narrow waterways connected to the ocean. For Marathon, although many damaged houses are near the coast, house type and house size are the two significant predictors (p-value < 0.001; Table 1b), highlighting the near-complete destruction of trailers (which are often small).

Possible measures to reduce flood vulnerability in the study areas include elevating and strengthening the buildings (especially mobile homes) and relocating homeowners living near the coastline (and narrow waterways) further inland. However, potential financial challenges exist, especially for Marathon, where the median annual income is $50,976 vs. $63,716 for Big Pine Key. Some local homeowners in a destroyed trailer community in Marathon (indicated by the red rectangle in Fig. 3d) with whom we talked had lived in trailers as their primary homes for decades without flood insurance. Financial constraints may hinder their rebuilding or relocating to somewhere safer. As low-income people living in mobile homes suffered most, natural hazards worsen economic inequality in this case. In contrast, discussion with local residents in Big Pine Key indicated that many structures there were second homes and, furthermore, were designed to withstand hurricane hazards (e.g., key assets were raised above the ground floor). These observations raise again issues of affordability and equity (Montgomery and Chakraborty, 2015). Policies relevant to hurricane damage recovery and rebuilding must address these issues.

Acknowledgments
This study is supported by NSF grant CMMI-1652448.

Marsooli, R., and N. Lin (2017). Numerical Modeling of Storm Tides and Waves Induced by Historical Tropical Cyclones along the East and Gulf Coasts of the United States. *Journal of Geophysical Research: Oceans*, Submitted.

Montgomery, M. C., & Chakraborty, J. (2015). Assessing the environmental justice consequences of flood risk: a case study in Miami, Florida. *Environmental Research Letters*, *10*(9), 095010.

Shao, W., Xian, S., Keim, B. D., Goidel, K., & Lin, N. (2017). Understanding perceptions of changing hurricane strength along the US Gulf coast. *International Journal of Climatology*, 37(4), 1716-1727.

Wang, C., Zhang, H., Feng, K., & Li, Q. (2017). Assessing hurricane damage costs in the presence of vulnerability model uncertainty. *Natural Hazards*, 85(3), 1621-1635.

Xian, S., Lin, N., & Hatzikyriakou, A. (2015). Storm surge damage to residential areas: a quantitative analysis for Hurricane Sandy in comparison with FEMA flood map. *Natural Hazards*, *79*(3), 1867-1888.

Xian, S., Lin, N., & Kunreuther, H. (2017). Optimal house elevation for reducing flood-related losses. *Journal of Hydrology*, *548*, 63-74.

Figures & Tables:

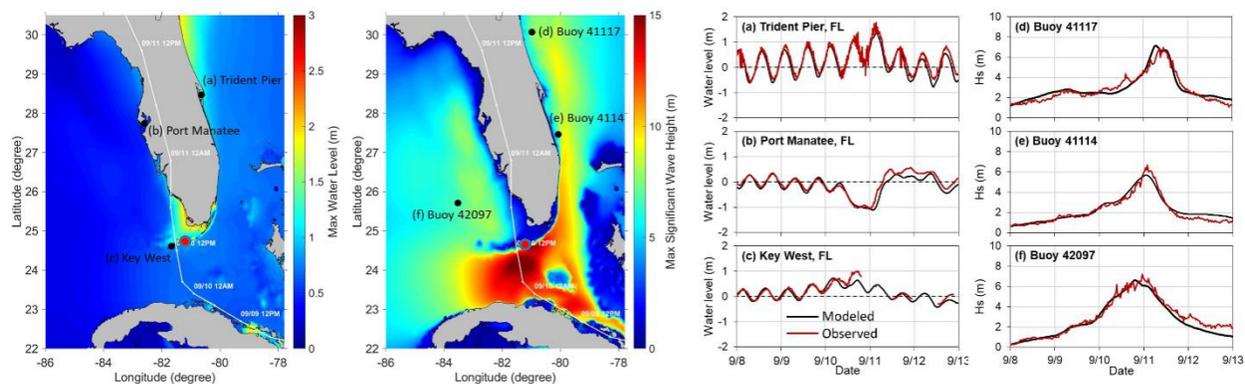

Figure 1. Hydrodynamic modeling of water level and wave height for Hurricane Irma. Left two panels show spatial distribution of modeled maximum water level and significant wave height, respectively. White curve represents storm track. Black points show locations of available tidal gauge and buoy stations. Red point indicates approximate location of study area. Right two panels compare observed and modeled time series of water level and significant wave height ($H_s$), respectively.

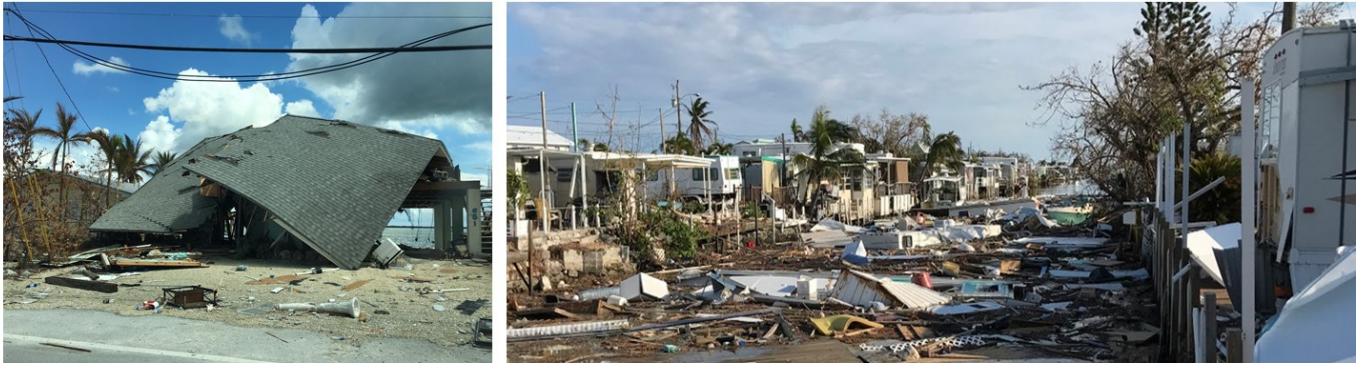

Figure 2. Photos of damage in (a) Big Pine Key: storm surge damage besides waterway (left side of building) and (b) Marathon: trailer community with house debris filling waterway

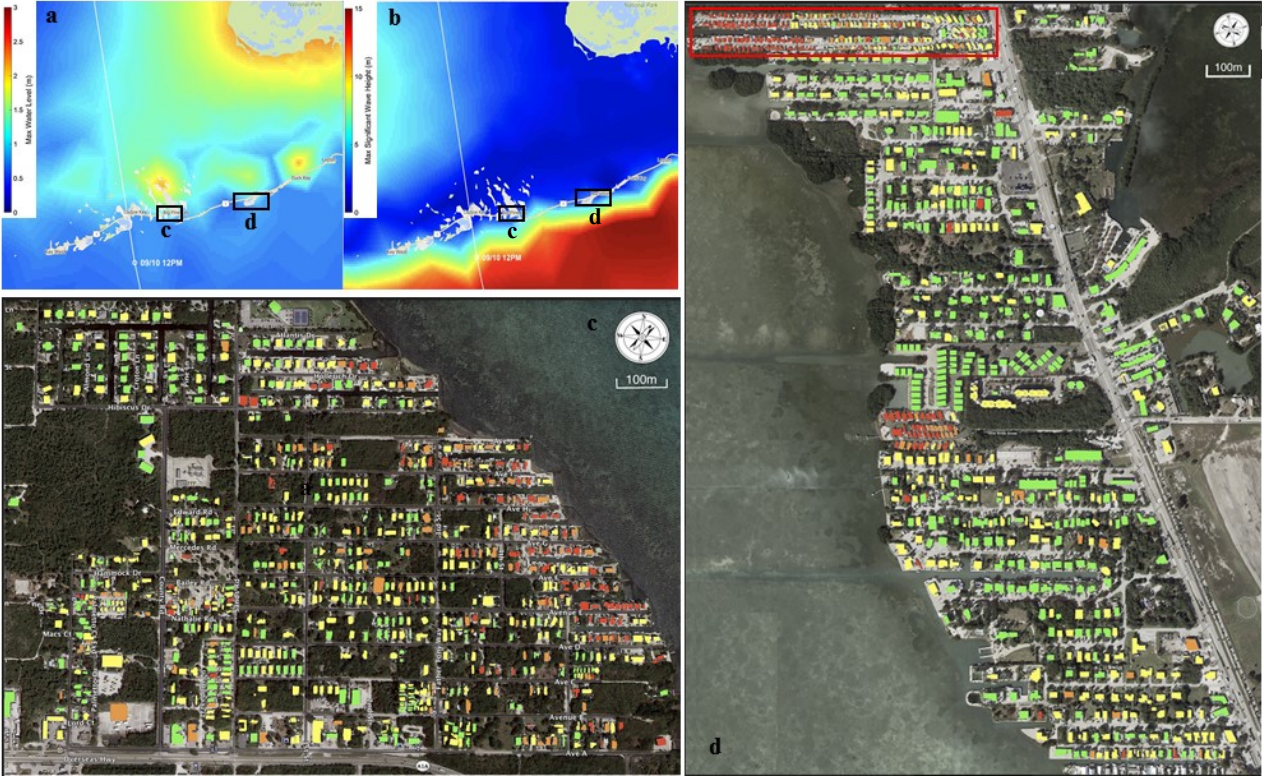

Figure 3. Spatial distribution of estimated hazards and damage states in study areas. (a) and (b) show simulated maximum total water level and significant wave height, respectively; (c) and (d) show assessed damage state (none: green; minor: yellow; major: orange; destroyed: red) for residential buildings in Big Pine Key and Marathon, respectively.

Table 1 Ordered logistic regression models that correlate damage state with vulnerability factors (a) for 846 assessed buildings in Big Pine Key; (b) for 811 buildings in Marathon.

| (a) Factors in damage state | Coef. | Std. Err. | z | p-value | 95% conf. interval | |
|---|---|---|---|---|---|---|
| House Type | 0.0233 | 1.987 | 0.12 | 0.906 | (-0.366 | 0.413) |

| | Coef. | Std. Err. | z | p-value | 95% conf. interval |
|---|---|---|---|---|---|
| House Size (square meters) | -0.00081 | 0.00059 | -1.36 | 0.174 | (-0.0198 0.000358) |
| Distance to Coast (meters) | 0.00718 | 0.00069 | 10.42 | 0.000 | (0.00583 0.00853) |

| (b) Factors in damage state | Coef. | Std. Err. | z | p-value | 95% conf. interval |
|---|---|---|---|---|---|
| House Type | -1.64 | 0.207 | -7.92 | 0.000 | (-2.05 - 1.236) |
| House Size (square meters) | -0.04961 | 0.001 | -4.88 | 0.000 | (-0.069 - 0.0029) |
| Distance to Coast (meters) | -0.0002145 | 0.00058 | -0.37 | 0.713 | (-0.0136 0.00093) |